\documentclass[aps,pre,reprint,             
superscriptaddress,amsmath,amssymb]{revtex4-2}

\usepackage{bm}
\usepackage{hyperref}
\usepackage{graphicx}
\usepackage{dcolumn}
\usepackage{amsmath}
\usepackage{mathrsfs}
\usepackage{subfigure}
\usepackage{enumerate}

\usepackage[dvips]{xcolor}

\newcommand{\df}[0]{D_{\rm f}}

\begin{document}

\title{Structural crossover of complex networks: bridging degree correlation and fractality}
\author{Shogo Mizutaka}
\email{shogo.mizutaka.kh50@vc.ibaraki.ac.jp}
\affiliation{Graduate School of Science and Engineering, Ibaraki University, Hitachi 316-8511, Japan}
\author{Jun Yamamoto}
\affiliation{Department of Network and Data Science, Central European University, Vienna 1100, Austria}
\author{Kousuke Yakubo}
\affiliation{National Institute of Technology, Asahikawa College, Asahikawa 071-8142, Japan}

\begin{abstract}
We investigate the relationship between long-range degree correlations and fractality in scale-free random networks. By analyzing degree correlations in scale-free random networks at the percolation threshold, at which a giant component emerges, we derive the ball volume $\tilde{\nu}_k(l)$, defined as the average number of nodes within distance $l$ from a root node with degree $k$. The resulting expression predicts a degree-dependent structural crossover. For $l\ll k^{1/(\df-1)}$, the ball volume grows as $\tilde{\nu}_k(l)\sim kl$, whereas for $l\gg k^{1/(\df-1)}$, it crosses over to the global fractal scaling $\tilde{\nu}_k(l)\sim l^{\df}$. Here, $\df$ is the fractal dimension of critical scale-free random networks. We further show that the structural crossover can be naturally understood in terms of critical branching processes and is also observed in empirical fractal networks.
\end{abstract}

\maketitle

\section{Introduction}

Network representations of complex systems in nature and society provide a powerful framework for understanding their structure and dynamics.
By representing system components as nodes and their interactions as edges, networks encode essential structural information in their topology.
One of the most common properties of real networks is degree heterogeneity.
In particular, empirical systems often exhibit a power-law tail in the degree distribution $p(k)\sim k^{-\gamma}$, which implies the coexistence of many low-degree nodes and a small number of hubs with extremely high degrees~\cite{Barabasi99,Albert02,Newman03a}.
Here, the degree exponent $\gamma$ characterizes the degree heterogeneity.
This heterogeneity profoundly influences a wide range of processes, such as percolation, epidemic spreading, and synchronization~\cite{Cohen00,Pastor01a,Ichinomiya04,Dorogovtsev08}.

In the presence of degree heterogeneity, correlations between degrees provide another layer of structural complexity.
The most commonly studied are nearest-neighbor degree correlations, characterized by the joint probability $p(k,k')$ that the two ends of a randomly chosen edge have degrees $k$ and $k'$.
Analyzing nearest-neighbor degree correlations reveals whether high-degree nodes tend to connect preferentially to other high-degree nodes (assortative mixing) or to low-degree ones (disassortative mixing)~\cite{Newman02,Newman03b}.
Social networks commonly exhibit assortative mixing, whereas technological and biological networks tend to show disassortative mixing.
These lowest-order, short-range degree correlations are known to affect several dynamical processes~\cite{Pastor01b,Maslov02,Gallos08}.
However, an accumulating body of evidence shows that real networks cannot be fully characterized by degree heterogeneity and nearest-neighbor degree correlations alone~\cite{Orsini15,Fujiki18}.
For instance, empirical systems such as the World Wide Web display a long-range hub-repulsive correlation~\cite{Song06,Fujiki17}.
The comparison of empirical networks with a null model constructed by randomly rewiring edges while preserving the joint probability $p(k,k^\prime)$ indicates the presence of organization patterns that extend beyond nearest neighbors~\cite{Fujiki17}.
This motivates the study of long-range degree correlations, which quantify the statistical dependence between degrees of nodes separated by more than one edge.
Long-range degree correlations have been investigated using various approaches, such as the assortativity coefficient between nodes at shortest-path distance $l$~\cite{Mayo15,Arcagni17}, analogies with long-range correlations in time series~\cite{Rybski10}, and characterizations in terms of generalized probability distributions of two node degrees and their shortest-path distance~\cite{Fujiki18,Fujiki20}.
While these studies have shown that long-range degree correlations are common features of empirical networks, their relationships with other structural properties and dynamical processes remain largely unresolved.

One structural property naturally associated with organization across multiple length scales is network fractality, which has been reported in empirical systems such as the World Wide Web and protein-protein interaction networks~\cite{Song05,Song06,Song07}.
Fractality is typically assessed in networks using a box-covering method: if the network is fractal, the minimum number $N(l)$ of boxes (subgraphs) of diameter $l$ required to cover the network follows a power law $N(l)\sim l^{-\df}$, where $\df$ denotes the fractal dimension~\cite{Song05,Song07}.
Correspondingly, the mean mass $\langle \nu(l) \rangle$ of a box of diameter $l$ is expected to scale as $\langle \nu(l) \rangle \sim l^{\df}$.
Such self-similar scaling indicates that the network exhibits nontrivial structural organization across a wide range of length scales.
From this viewpoint, fractality in networks can be interpreted as a manifestation of correlations that extend over long distances.
It is therefore natural to expect that long-range degree correlations and network fractality are closely related, and clarifying this relationship is crucial for understanding the large-scale architecture of real-world networks.

To gain theoretical insight into this relationship, random networks offer a useful and analytically tractable framework.
Random networks undergo a percolation transition at a critical point above which a giant component emerges.
Near the percolation critical point, the system is characterized only by a correlation length $\xi_l$: within distances $l \lesssim \xi_l$, nontrivial degree correlations persist, whereas correlations decay beyond the scale $\xi_l$~\cite{Cohen04,Mizutaka20}.
At criticality, where $\xi_l$ diverges, the degree correlations extend throughout the incipient giant component (critical component), and the component becomes globally fractal.
For scale-free random networks, the fractal dimension of the critical component has been shown to depend on the degree exponent $\gamma$:
$\df = (\gamma - 2)/(\gamma - 3) \ \text{for} \ 3 < \gamma < 4 \ \text{and} \ \df = 2 \ \text{for} \ \gamma \ge 4$~\cite{Bruda01,Cohen03,Cohen04}.
Despite these advances, relatively little is known about the connection between fractality and long-range degree correlations in random networks, as the two properties have been studied separately.

In this work, we analytically relate long-range degree correlations to fractality in critical scale-free random networks and reveal that this relationship is also observed in backbone structures of empirical fractal networks.
Our focus is on the conditional probability that a node separated by $l$ from a node (root node) of degree $k$ has degree $k'$. From this probability, we derive the ball volume $\tilde{\nu}_k(l)$, the expected number of nodes within distance $l$ from a root node of degree $k$ (excluding the root), as
\begin{align}
	\tilde{\nu}_k(l)=kl+cl^{\df-1}(l-1),
	\label{eq:nu_tilde}
\end{align}
where $c$ is a constant.
The result in Eq.~(\ref{eq:nu_tilde}) recovers the known fractal dimension $\df$ of critical scale-free random networks in the limit $l\to \infty$~\cite{Bruda01,Cohen03,Cohen04}.
More importantly, Eq.~(\ref{eq:nu_tilde}) predicts a degree-dependent structural crossover at $l_{\rm co}(k)\sim k^{1/(\df-1)}$: the volume $\tilde{\nu}_k(l)$ scales as $\tilde{\nu}_k(l)\sim kl$ for $l\ll l_{\rm co}(k)$, whereas it scales as $\tilde{\nu}_k(l)\sim l^{\df}$ for $l\gg l_{\rm co}(k)$.
Thus, critical scale-free random networks exhibit a local regime with linear growth, $\tilde{\nu}_k(l)\sim kl$, and a global fractal regime, $\tilde{\nu}_k(l)\sim l^{\df}$.
In contrast to previous studies that characterize fractal structure in the limit $l\to \infty$, we demonstrate that long-range degree correlations encode a degree-dependent structural crossover.
Furthermore, we argue that a critical branching process provides a physical interpretation of the structural crossover predicted from long-range degree correlations. 
This interpretation links the structural crossover to critical branching rather than to a feature specific to random-network construction.
Based on this finding, we analyze empirical fractal networks and reveal the existence of similar linear growth laws.

The rest of this paper is organized as follows. 
Section~II develops our theory for scale-free random networks and derives the main result, Eq.~(\ref{eq:nu_tilde}).
In Sec.~III, we discuss a branching-process interpretation of the crossover and its relation to previous work and address the implications for real-world networks. 
Section~IV summarizes and concludes the paper with an outlook.

\section{Structural Crossover in Critical Random Networks}

To derive Eq.~(\ref{eq:nu_tilde}), we consider probability distributions defined on the giant components of scale-free random networks. 
We first briefly review these distributions and examine their critical limits. 
Next, we obtain the ball volume $\tilde{\nu}_k(l)$ from a cluster-growing analysis based on a probability distribution describing long-range degree correlations. 
Finally, we will numerically verify Eq.~(\ref{eq:nu_tilde}) using a scaling argument.

Let us consider the configuration model, which generates sparse, locally tree-like, and degree-uncorrelated networks with degree distribution $p(k)$, and assume that the generated network contains a giant component (GC).
We denote by $u$ the probability that a randomly chosen edge does not lead to the GC.
The probability that a randomly chosen node has degree $k$ and belongs to the GC is $(1-u^k)p(k)$.
The summation of the probability over degree $k$ corresponds to the relative size $S$ of the GC: $S=\sum_k (1-u^k)p(k)$.
Thus, the degree distribution $p_{\rm GC}(k)$ in the GC is
\begin{align}
    p_{\rm GC}(k)=\frac{(1-u^k)p(k)}{S}.
    \label{eq:pk_gc}
\end{align}
Similarly, a randomly chosen edge belongs to the GC with probability $1-u^2$, while an edge whose endpoints have degrees $k$ and $k'$ belongs to the GC with probability $1-u^{k+k'-2}$. 
Therefore, the endpoint-degree distribution of edges in the GC is given by
\begin{align}
    p_{\rm GC}(k,k')=\frac{1-u^{k+k'-2}}{1-u^2}q(k)q(k'),
\end{align}
where $q(k)$ is the neighbor degree distribution $q(k)=kp(k)/z_1$ and $z_1$ is the first moment of $p(k)$ (the average degree).
Extending this argument based on the probability $u$, we can also analyze the conditional probability $p_{\rm GC}(k'|k,l)$ that a node separated by distance $l$ from a degree-$k$ node has degree $k'$ on the GC:
\begin{align}
	p_{\rm GC}(k'|k,l)=\frac{1-u^{k+k'-2} v^{l-1}}{1-u^k v^{l-1}}q(k').
	\label{eq:pkkl_gc}
\end{align}
Here, the variable $v$ is defined by $v=G'_1(u)/G'_1(1)$ and controls the decay rate of distance-dependent correlations, where the function $G_1(u)=\sum_k q(k)u^{k-1}$ is the generating function associated with neighbor-degree distribution $q(k)$. The variable $u$ is determined by the self-consistent equation $u=G_1(u)$.

We now consider the critical limit, in which the GC is globally fractal.
For an infinitesimal $\epsilon=1-u$, $v=1-z_{3/2}\epsilon$.
Here, $z_{3/2}\equiv z_3/z_2$ is the ratio of the factorial moments $z_3=\sum_k k(k-1)(k-2)p(k)$ and $z_2=\sum_k k(k-1)p(k)$ of the degree distribution $p(k)$.
Substituting $u=1-\epsilon$ and $v=1-z_{3/2}\epsilon$ into Eqs.~(\ref{eq:pk_gc}) and (\ref{eq:pkkl_gc}) and taking the limit $\epsilon\to 0$ yields the following expressions for degree distribution $\tilde{p}(k)$ and conditional probability $\tilde{p}(k'| k,l)$ on the critical component: 
\begin{align}
	\tilde{p}(k)=\frac{kp(k)}{z_1}=q(k)
	\label{eq:critical_pk}
\end{align}
and
\begin{align}
	\tilde{p}(k'|k,l)=\frac{(k+k'-2)+z_{3/2}(l-1)}{k+z_{3/2}(l-1)}q(k'),
	\label{eq:critical_pkkl}
\end{align}
respectively \cite{Mizutaka20,Dorogovtsev08}. 
In what follows, quantities associated with the GC at criticality are denoted by symbols with tilde.
Using the degree distribution (\ref{eq:critical_pk}) and the Molloy-Reed criterion $z_{2}/z_{1}=1$ for criticality \cite{Molloy95},
we can rewrite $z_{3/2}$ as
\begin{align}
	z_{3/2}=\sum_k (k-1)(k-2)\tilde{p}(k).
\end{align}

Next, we connect the long-range degree correlations and fractality of the critical component by analyzing Eq.~(\ref{eq:critical_pkkl}).
Let $\tilde{s}(k'|k,l)$ be the number of nodes of degree $k'$ at distance $l$ from a node of degree $k$. Then, we have
\begin{align}
    \tilde{s}(k'|k,l)=\tilde{p}(k'|k,l)\sum_{k''}(k''-1)\tilde{s}(k''|k,l-1).
\end{align}
From the initial condition $\tilde{s}(k'|k,l=1)=k\tilde{p}(k'|k,l=1)$ and Eq.~(\ref{eq:critical_pkkl}), induction gives
\begin{align}
    \tilde{s}(k'|k,l)=[(k+k'-2)+z_{3/2}(l-1)]q(k').
    \label{eq:ball_surface}
\end{align}
Thus, the shell size (surface area) $\tilde{s}_k(l)\equiv\sum_{k'}\tilde{s}(k'|k,l)$, i.e., the number of nodes at distance $l$ from a degree-$k$ node, is given by
\begin{align}
    \tilde{s}_k(l)=k+z_{3/2}(l-1).
    \label{eq:ball_surface_explicit}
\end{align}
This yields the volume $\tilde{\nu}_k(l)$ of a ball of radius $l$ centered at a degree-$k$ root node, excluding the root, as
\begin{align}
	\tilde{\nu}_k(l)=kl+\frac{z_{3/2}}{2}l(l-1),
	 \label{eq:volume_form_k_gen}
\end{align}
by summing the shell sizes $\tilde{s}_{k}(l)$ over distance $l$ from $l=1$. 
Since the ball volume differs from $\tilde{\nu}_{k}(l)$ by only one, we hereafter treat $\tilde{\nu}_{k}(l)$ as the ball volume.

The behavior of $z_{3/2}$ in Eq.~(\ref{eq:volume_form_k_gen}) depends on the value of the degree exponent $\gamma$ of critical scale-free random networks. 
In the infinite-system limit, both $z_{2}$ and $z_{3}$ converge to finite values with $z_{2}\neq0$ when $\gamma>4$, so that $z_{3/2}$ is a finite constant. 
Thus, Eq.~(\ref{eq:volume_form_k_gen}) provides an exact expression for the mean-field cases such as the Erd\H{o}s-R\'{e}nyi random graph (ERRG). 
In contrast, when $3<\gamma\le 4$, $z_{3}$ (and thus $z_{3/2}$) diverges in the infinite-system limit. 
However, if we measure the volume from a root node with finite degree, the volume within any finite radius remains finite almost surely because infinite hubs have probability measure zero. 
Assuming Eq.~(\ref{eq:volume_form_k_gen}) remains valid in this situation, we estimate the effective value of $z_{3/2}$ within a ball of radius $l$ from its maximum degree $\tilde{k}_{\rm c}(l)$. 
This effective value diverges as the maximum degree increases.
Estimating $z_{3/2}$ from 
\begin{align}
    z_{3}&=\sum_{k}^{\tilde{k}_{\rm c}(l)}(k-1)(k-2)\tilde{p}(k)\nonumber \\
    &\sim \int_{}^{\tilde{k}_{\rm c}(l)}(k-1)(k-2)\tilde{p}(k)dk,
\end{align}
we obtain
\begin{align}
    z_{3/2}(l) \sim [\tilde{k}_{\rm c}(l)]^{4-\gamma}
    \label{eq:z32}
\end{align}
for $3<\gamma<4$.
The maximum degree $\tilde{k}_{\rm c}(l)$ is then estimated using the natural-cutoff condition: $1/\tilde{\nu}_k (l)\sim \sum_{k=\tilde{k}_{\rm c}(l)}^{\infty} \tilde{p}(k)$, so that the expected number of nodes with degree greater than $\tilde{k}_{\rm c}$ is of order unity.
From Eq.~(\ref{eq:critical_pk}), we find the relation between the maximum degree $\tilde{k}_{\rm c}(l)$ and the ball volume $\tilde{\nu}_k(l)$ as
\begin{align}
    \tilde{k}_{\rm c}(l) \sim \tilde{\nu}_k(l)^{1/(\gamma-2)}.
    \label{eq:natural_cutoff}
\end{align}
For sufficiently large $l$, Eq.~(\ref{eq:volume_form_k_gen}) behaves as $\tilde{\nu}_k(l)\sim z_{3/2}(l)\,l^2$. Substituting Eqs.~(\ref{eq:z32}) and (\ref{eq:natural_cutoff}) into this relation and solving for $\tilde{\nu}_{k}(l)$ yields
\begin{align}
    \tilde{\nu}_{k}(l) \sim l^{(\gamma-2)/(\gamma-3)}.
    \label{eq:ball_volume_gamma<4}
\end{align}
Here, the exponent $(\gamma-2)/(\gamma-3)$ is nothing but the fractal dimension $\df$ of scale-free random networks for $3<\gamma<4$.
By contrast, for $\gamma>4$, $z_{3/2}$ is always finite and we have $\tilde{\nu}_{k}(l)\sim l^{2}$. 
The exponent $2$ in this relation is also consistent with the fractal dimension $D_{\rm f}=2$ in the mean-field regime ($\gamma>4$). 
Combining these behaviors of $\tilde{\nu}_{k}(l)$, we find that Eq.~(\ref{eq:volume_form_k_gen}) reduces to Eq.~(\ref{eq:nu_tilde}), namely,
\begin{align}
	\tilde{\nu}_k(l)=kl+cl^{\df-1}(l-1). \nonumber
\end{align}
\begin{figure}[t!]
	\begin{center}
    \includegraphics[width=\linewidth]{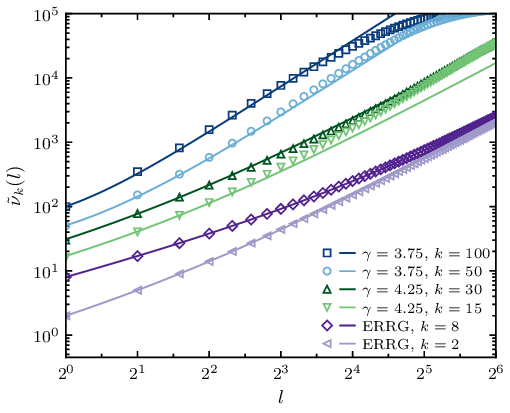}	
    \caption{
        Distance dependence of the ball volume $\tilde{\nu}_k(l)$.
        The abscissa $l$ represents the distance from a root node of degree $k$.
        Symbols show simulation results for critical scale-free random networks with $\gamma=3.75$ and $4.25$ and for Erd\H{o}s-R\'{e}nyi random graphs.
        Lines represent the theoretical prediction $\tilde{\nu}_k(l)=kl+c\,l^{\df-1}(l-1)$ with $\df=(\gamma-2)/(\gamma-3)$ for $3<\gamma<4$ and $\df=2$ for $\gamma\ge4$.
        In the numerical calculation, we generated $10^3$ networks with $10^7$ nodes.
        For scale-free networks with $\gamma=4.25$ and Erd\H{o}s-R\'{e}nyi random graphs, we sampled a node with degree $k$ from each network.
        For $\gamma=3.75$, we sampled a node with degree in $[k-2, k+2]$ from each network.
    }
   \label{fig:1}
	\end{center}
\end{figure}
The asymptotic form of this expression suggests a structural crossover depending on the distance $l$ from the root node:
\begin{align}
	\tilde{\nu}_k(l)\sim
    \begin{cases}
      k l  & (l\ll l_{\rm co}(k)) \\[8pt]
      l^{\df}   & (l\gg l_{\rm co}(k)).
    \end{cases}
\end{align}
Here, the degree-dependent crossover length $l_{\rm co}(k)$ is given by
\begin{align}
    l_{\rm co}(k) \sim (k/c)^{1/(\df-1)}.
    \label{eq:crossover-hetero}
\end{align}
Therefore, the volume of a ball centered at a degree-$k$ node behaves as $kl$ for $l\ll l_{\rm co}(k)$ and $l^{\df}$ for $l\gg l_{\rm co}(k)$, and the crossover length $l_{\rm co}(k)$ depends on the degree of the root node.
The long-range degree correlation [Eq.~\eqref{eq:critical_pkkl}] therefore contains information about both global fractal scaling and the degree-dependent crossover.

Figure~\ref{fig:1} shows the numerical results for the $l$-dependence of the ball volume $\tilde{\nu}_k(l)$.
We plot the results for scale-free networks with two representative degree exponents, $\gamma=3.75$ and $\gamma=4.25$ at criticality and for the ERRG with the average degree $z_1 = 1$.
Here, the scale-free networks are generated from the configuration model whose degree distribution obeys
\begin{align}
	 p(k)=
	 \begin{cases}
    1-r & (k=1) \\[4pt]
    \mathcal{C} r k^{-\gamma} &(k>1),
    \end{cases}
  \label{eq:pk}
\end{align}
where $\mathcal{C}$ is the normalization constant.
We determine the control parameter $r$ to satisfy the Molloy-Reed criterion $z_2/z_1=1$ at criticality.
Symbols in Fig.~\ref{fig:1} represent numerical results for several root degrees ($k=50,\,100$ for $\gamma=3.75$, $k=15,\,30$ for $\gamma=4.25$, and $k=2,\,8$ for ERRG), while the lines correspond to the analytical prediction given by Eq.~(\ref{eq:nu_tilde}). 
Here, the coefficient $c$ in Eq.~(\ref{eq:nu_tilde}) is estimated from the plateau value of $c=[\tilde{\nu}_k(l)-kl]/[l^{\df-1}(l-1)]$ for scale-free networks, while we fit the exact expression $\tilde{\nu}_k(l)=kl+l^{\df-1}(l-1)/2$ for ERRG. See the Appendix for the estimation of $c$, the results for ERRG, and a decomposition into the two terms $kl$ and $l^{D_{\rm f}}$.
The agreement between numerical results and theory demonstrates that Eq.~(\ref{eq:nu_tilde}) correctly captures the distance dependence of the volume across distinct parameter sets.
Namely, at short distances, the volume grows linearly as $\tilde{\nu}_k(l)\sim kl$, whereas at larger distances the growth crosses over to a well-known fractal regime, $\tilde{\nu}_k(l)\sim l^{\df}$.

While Fig.~\ref{fig:1} confirms Eq.~(\ref{eq:nu_tilde}) for different pairs of $(\gamma,k)$, the crossover behavior is not clearly visible in the raw $\tilde{\nu}_k(l)$ plot.
To make the structural crossover clear, we examine the following rescaled volume $F(l)$:
\begin{align}
    F(l) \equiv \frac{\tilde{\nu}_k(l)}{cl^{\df-1}(l-1)}.	
    \label{eq:F_l}
\end{align}
Substituting Eq.~(\ref{eq:nu_tilde}) into Eq.~(\ref{eq:F_l}) and using Eq.~(\ref{eq:crossover-hetero}), we obtain $F[l(x)]=1+[l/(l-1)]\,x^{1-\df}\approx 1+x^{1-\df}$ for large $l$, where $x \equiv l/l_{\mathrm{co}}(k)$ is the rescaled distance.
This predicts the behavior
\begin{align}
	F[l(x)]\sim
    \begin{cases}
    x^{1-\df} & (x\ll 1) \\
    1 & (x\gg 1).
    \end{cases}
    \label{eq:crossover-fx}
\end{align}
\begin{figure}[t!]
	\begin{center}
    \includegraphics[width=\linewidth]{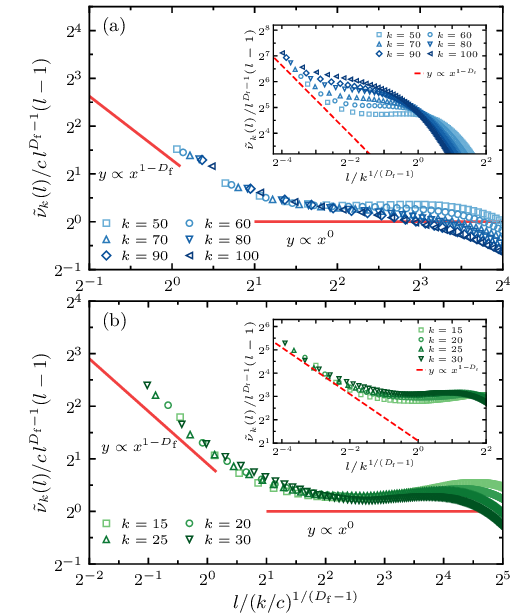}	
    \caption{
        Scaling collapse of the rescaled ball volume, demonstrating the degree-dependent crossover predicted by Eq.~(\ref{eq:crossover-hetero}).
        Rescaled ball volume $F(l)\equiv \tilde{\nu}_k(l)/[cl^{\df-1}(l-1)]$ is plotted as a function of the rescaled distance $x\equiv l/l_{\mathrm{co}}(k)$ with $l_{\mathrm{co}}(k)=(k/c)^{1/(\df-1)}$ for (a) scale-free networks with $\gamma=3.75$ and (b) those with $\gamma=4.25$. Symbols correspond to different root degrees $k$.
        Red dashed lines indicate the theoretical behaviors $F[l(x)]\propto x^{1-\df}$ for $x\ll 1$ and $F[l(x)]\propto x^{0}$ for $x\gg 1$.
        In the insets of both panels, the rescaled ball volume is plotted as a function of $l/k^{1/(\df-1)}$, which enables us to verify Eq.~(\ref{eq:crossover-hetero}) without estimating the coefficient $c$.
        All data are obtained under the same conditions as in Fig.~\ref{fig:1}.
    }
     \label{fig:2}
	\end{center}
\end{figure}
Figures~\ref{fig:2} (a) and (b) test the above scaling form for $\gamma=3.75$ and $\gamma=4.25$, respectively.
The numerical results for various values of the degree $k$ collapse onto a common  curve.
In the small-$x$ regime, the results are consistent with the predicted power law $F[l(x)]\sim x^{1-\df}$, and $F[l(x)]$ approaches $1$ for $x\gg 1$.
For $x\gg 2^{4}$, the curves bend downward from the plateau due to the finite-size effect.
In the insets of both panels, we verify the validity of Eq.~(\ref{eq:crossover-hetero}) without estimating the coefficient $c$.
Since
\begin{align}
    \frac{\tilde{\nu}_{k}(l)}{l^{D_{\text{f}}-1}(l-1)}\simeq \left[\frac{l}{k^{1/(D_{\rm f}-1)}}\right]^{1-D_{\text{f}}}+c
    \label{eq:Fl_wo_c}
\end{align}
holds, the left-hand side of Eq.~(\ref{eq:Fl_wo_c}) is plotted, i.e., $\tilde{\nu}_k(l)/[l^{\df-1}(l-1)]$ as a function of $l/k^{1/(\df-1)}$. 
As a result, the first term on the right-hand side of Eq.~(\ref{eq:Fl_wo_c}) is observed for small values of $l/k^{1/(D_{\text{f}}-1)}$, while the result approaches a constant $c$ for large values.
Our numerical results shown in Figs.~\ref{fig:1} and \ref{fig:2} strongly support the validity of the theoretical results Eqs.~(\ref{eq:nu_tilde}) and (\ref{eq:crossover-hetero}). 
We have also confirmed the scaling collapse for ERRG, as shown in Appendix A.
While several studies have reported the fractal behavior of scale-free random networks for $l\gg 1$~\cite{Bruda01,Cohen03, Cohen04}, the present work additionally predicts the linear growth behavior for $l \ll l_{\rm co}(k)$.
It should be noted that the term $kl$ originates from a globally correlated structure of the critical component in a scale-free random network.

\section{Branching Process Interpretation}
While long-range degree correlations predict detailed structures of critical random networks, it is unclear why the chain-like structure characterized by $\tilde{\nu}_{k}(l)\sim kl$ appears in addition to the fractal structures, for which $\tilde{\nu}_{k}(l)\sim l^{\df}$ holds. 
We propose a physical explanation for the origin of the chain-like structure via a critical branching process.
To this end, let us recall the direct calculation of fractal dimensions using the critical branching process proposed by Cohen and Havlin~\cite{Cohen04}.
In Ref.~\cite{Cohen04}, the authors have shown that the survival probability $\rho_{l}$ of a branch up to distance $l$ behaves as $\rho_{l} \sim l^{-(\df-1)}$. 
Accordingly, $\rho_{l}^{-1}\sim l^{\df-1}$ gives the mean size of the surviving $l$-th shell.
Consider the case of growing a ball from a root node with degree $k$.
The probability that at least one of the $k$ branches emanating from the root survives up to distance $l$ is $1-(1-\rho_{l})^k$.
In critical branching processes, each branch has one offspring on average.
The mean shell size at distance $l$, conditioned on the survival of at least one branch, is therefore
\begin{align}
    \tilde{s}_k(l)=\frac{k}{1-(1-\rho_{l})^k}.
    \label{eq:cohen-havlin}
\end{align}
This expression has two asymptotic regimes.
Since $\rho_{l}\sim l^{-(\df-1)}$, the denominator of Eq.~(\ref{eq:cohen-havlin}) is approximated by $1-(1-\rho_{l})^{k}\approx 1$ for $l\ll k^{1/(\df-1)}$.
Thus, in this regime, Eq.~(\ref{eq:cohen-havlin}) behaves as $\tilde{s}_k(l)\sim k$, indicating $\tilde{\nu}_k(l)\sim kl$ ($kl$ regime).
By contrast, when $l\gg k^{1/(\df-1)}$ so that $\rho_{l}\ll 1$, the denominator $1-(1-\rho_{l})^{k}$ is approximated by $k\rho_{l}$.
In this regime, Eq.~(\ref{eq:cohen-havlin}) behaves as $\tilde{s}_k(l)\sim l^{\df-1}$, indicating $\tilde{\nu}_k(l)\sim l^{\df}$ (fractal regime).
The crossover between the two regimes is determined by $l_{\rm co}(k) \sim k^{1/(\df-1)}$.
Thus, the distance dependence of the ball volume and the degree-dependent crossover [Eqs.~(\ref{eq:nu_tilde}) and (\ref{eq:crossover-hetero})] obtained from the argument of the long-range degree correlations has an interpretation in terms of a critical branching process.
This suggests that the structural crossover might not be intrinsic to random networks but rather related to critical branching processes. 
The condition $k\rho_{l}\sim 1$ separates two regimes in critical branching processes, where $k\rho_{l}$ is the expected number of surviving branches among the $k$ branches. 
Thus, the crossover length $l_{\rm co}(k)$ represents the characteristic distance at which the expected number of surviving branches becomes of order unity.

\begin{figure}[tb]
	\begin{center}
    \includegraphics[width=1.\linewidth]{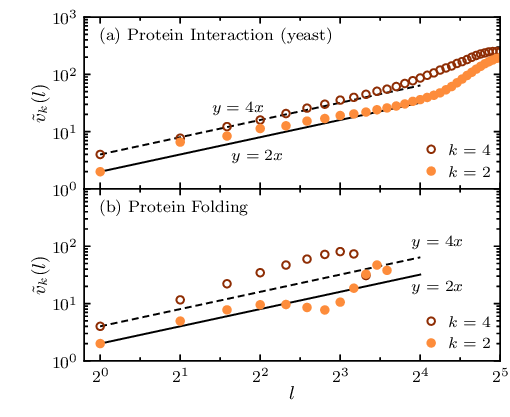}	
    \caption{
        Residual ball volume $\tilde{v}_k(l)\equiv\tilde{\nu}_k(l)-c l^{\df-1}(l-1)$ as a function of distance $l$ for the skeletons of (a) a protein interaction network of yeast and (b) a protein-folding network. 
        Here, $\tilde{\nu}_k(l)$ is  the measured average volume from a root node of degree $k$, and $\df$ and $c$ are estimated from the data. 
        In both panels, open and filled symbols correspond to $k=4$ and $k=2$, respectively. 
        The dashed line is a visual guide.
    }
    \label{fig:3}
	\end{center}
\end{figure}

Since we attribute the origin of the structural crossover to critical branching processes, it is natural to expect that a similar behavior of $\tilde{\nu}_k(l)$ is observed in the skeletons of empirical fractal networks, which are reported to possess properties of critical branching trees~\cite{Goh06,Kim07}.
Here, the skeleton is an underlying backbone structure that constitutes a spanning tree formed by edges with high betweenness centralities~\cite{Goh06,Kim07}.
In empirical tests, we confirm that $\tilde{\nu}_{k}(l)$ contains the $kl$ term in addition to the fractal scaling term, as predicted by Eq.~(\ref{eq:nu_tilde}). 
Figure~\ref{fig:3} shows the distance dependence of the residual volume $\tilde{v}_{k}(l)\equiv \tilde{\nu}_{k}(l)-cl^{\df-1}(l-1)$ for two real-world fractal networks,
where $\tilde{\nu}_k(l)$ is the observable measured from a root node of degree $k$, and $\df$ and $c$ are estimated numerically (see Appendix B).
Panels (a) and (b) indicate the results for the skeletons of a protein interaction network~\cite{Jeong01,Coulomb05} and a protein folding network~\cite{Bu03}, respectively, which are known to be fractal.
In both networks, the residual volume $\tilde{v}_{k}(l)$ increases approximately linearly with $l$ over a certain range and the proportionality coefficient is given by $k$.
This implies that the $kl$ term contributes to the ball volume $\tilde{\nu}_{k}(l)$, as shown in Eq.~(\ref{eq:nu_tilde}). 
These results demonstrate that the backbone structures of real-world fractal networks exhibit degree-dependent chain-like structures at shorter length scales, in addition to the fractal behavior observed at longer length scales.
The deviation from linearity in $\tilde{v}_{k}(l)$ suggests that loops and shortcuts in the original networks may obscure the chain-like contribution carried by their skeletons.
If the backbone structure of an empirical fractal network is represented by a critical branching tree, as proposed previously in~\cite{Goh06,Kim07}, the present analysis provides a quantitative description of how a local chain-like structure crosses over to a global fractal structure. 
Our analysis further shows that the critical branching trees and skeletons of empirical fractal networks share not only fractal properties but also a similar linear structure at shorter length scales.

\section{Summary and Conclusion}

In this work, we have studied the relationship between long-range degree correlations and fractality in scale-free random networks. 
By focusing on the conditional probability that a node separated by distance $l$ from a degree-$k$ node has degree $k'$, we derived the ball volume $\tilde{\nu}_k(l)$ around a degree-$k$ root node at criticality: $\tilde{\nu}_k(l)=kl+c l^{\df-1}(l-1)$.
This shows that the volume growth can be decomposed into two contributions. 
The first term $kl$ represents the degree-dependent linear growth from the root node, while the second term represents the known fractal growth characterized by the fractal dimension $\df$. 
An important consequence of this expression is the degree-dependent structural crossover. 
For $l\ll l_{\rm co}(k)$, the ball volume grows as $\tilde{\nu}_k(l)\sim kl$, whereas for $l\gg l_{\rm co}(k)$ it obeys the fractal scaling $\tilde{\nu}_k(l)\sim l^{\df}$. 
Here, $l_{\rm co}(k)\sim k^{1/(\df-1)}$ is the crossover length, which depends on the degree of a root node.
We have confirmed this theoretical prediction by numerical simulations of critical scale-free random networks. 
In both the regime $3<\gamma<4$, where $\df=(\gamma-2)/(\gamma-3)$, and the mean-field regime $\gamma\ge4$, where $\df=2$, the measured ball volume agrees well with the theoretical form. 
By applying a scaling analysis to $\tilde{\nu}_k(l)$, we numerically confirmed the theoretical prediction for the degree dependence of the crossover length $l_{\rm co}(k)$, namely, $l_{\rm co}(k)\sim k^{1/(\df-1)}$.
These numerical results support the coexistence of the linear contribution $kl$ and the fractal contribution $l^{\df}$ to the ball volume $\tilde{\nu}_k(l)$, in agreement with the proposed crossover form.
We have further provided a physical interpretation of the structural crossover in terms of critical branching processes. Consistent with this interpretation, the backbones extracted from empirical fractal networks exhibit a similar degree-dependent linear-growth regime, which suggests the backbones (skeletons) of them possess critical-branching-tree-like structures.

The present study raises several questions for future work.
We demonstrate a structural crossover in scale-free random networks with fractality and find a similar signature in two empirical networks.
A fundamental question is whether the crossover predicted in this study is a universal property of fractal scale-free networks, independent of the specific network type.
Likewise, it is important to clarify the relationship between the structural crossover found here and dynamical processes such as random walks and transport.
Since these processes are sensitive to both degree heterogeneity and global fractal structure, the degree-dependent crossover may provide a useful structural basis for understanding how their dynamics change across the two structural regimes.

\begin{acknowledgements}
S.M.\ was supported by JSPS (Japan) KAKENHI (Grant Number 24K06896) and by JST ERATO Grant Number JPMJER2401.
J.Y.\ acknowledges support from ERC Grant No.~810115 (DYNASNET) and from the National Science Foundation through the AccelNet-MultiNet program under Awards Nos.~1927425 and 1927418.
\end{acknowledgements}

\clearpage
\onecolumngrid
\appendix

\section{Erd\H{o}s-R\'{e}nyi random graphs}

\begin{figure}[t!]
	\begin{center}
		\includegraphics[width=1\linewidth]{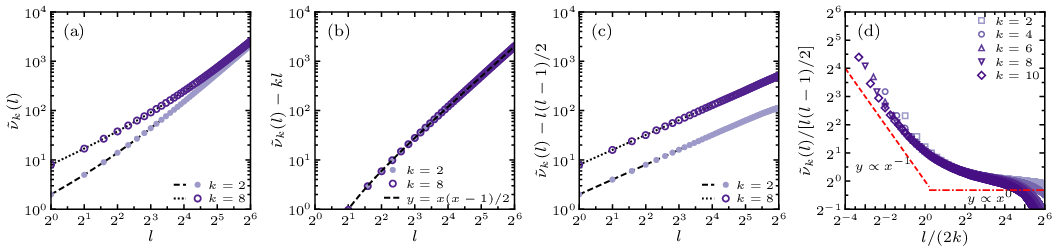}
        \caption{
            Decomposition of the ball volume in critical Erd\H{o}s--R\'enyi random graphs with mean degree $z_1=1$.
            (a) Ball volume $\tilde{\nu}_k(l)$ as a function of distance $l$ for root node degrees $k=2$ (filled circles) and $k=8$ (open circles). The dashed and dotted lines show the exact expression $\tilde{\nu}_k(l)=kl+l(l-1)/2$ for $k=2$ and $8$, respectively.
            (b) Fractal contribution obtained by subtracting the linear term, $\tilde{\nu}_k(l)-kl$. Data for both root degrees collapse onto the exact result $l(l-1)/2$ (dashed line).
            (c) Linear contribution obtained by subtracting the fractal term, $\tilde{\nu}_k(l)-l(l-1)/2$. The dashed and dotted lines represent $kl$ for $k=2$ and $8$, respectively.
            (d) Scaling collapse of $\tilde{\nu}_k(l)/[l(l-1)/2]$ as a function of $l/(2k)$ for $k=2$, $4$, $6$, $8$, and $10$. The red dashed guides indicate the asymptotic behaviors proportional to $x^{-1}$ for $x\ll1$ and $x^0$ for $x\gg1$, where $x=l/(2k)$.
        }
   \label{fig:app-errg}
	\end{center}
\end{figure}

We derive the ball volume for critical Erd\H{o}s--R\'enyi random graphs by evaluating the factorial moments $z_2$ and $z_3$ at finite system size and then taking the thermodynamic limit.
Consider a graph with $N$ nodes in which each pair of distinct nodes is independently connected with probability $\rho$.
The degree distribution is
\begin{equation*}
    p_N(k)=\binom{N-1}{k}\rho^k(1-\rho)^{N-1-k},
\end{equation*}
and the mean degree is $z_1=(N-1)\rho$.
The corresponding generating function,
$G_N(x)=\sum_k p_N(k)x^k=(1-\rho+\rho x)^{N-1}$,
gives the second and third factorial moments as
\begin{align*}
    z_2&=G_N''(1)=\frac{N-2}{N-1}z_1^2,
    \qquad
    z_3=G_N'''(1)=\frac{(N-2)(N-3)}{(N-1)^2}z_1^3.
\end{align*}
Hence,
\begin{equation*}
    \frac{z_2}{z_1}=\frac{N-2}{N-1}z_1,
    \qquad
    \frac{z_3}{z_2}=\frac{N-3}{N-1}z_1.
\end{equation*}
In the thermodynamic limit ($N\to \infty$), therefore, the Molloy-Reed criterion $z_2/z_1=1$ reduces to $z_1=1$, which gives
\begin{equation*}
z_{3/2}=\frac{z_3}{z_2}=1.
\end{equation*}
Substituting this result into Eq.~(\ref{eq:volume_form_k_gen}) gives the
exact expression in this limit,
\begin{equation}
    \tilde{\nu}_k(l)
    =kl+\frac{1}{2}l(l-1).
    \label{eq:ball_volume_errg}
\end{equation}

The numerical results in Fig.~\ref{fig:app-errg}(a) agree well with Eq.~(\ref{eq:ball_volume_errg}). 
Figures~\ref{fig:app-errg}(b) and (c) isolate the root-degree-independent quadratic contribution and the root-degree-dependent linear contribution, respectively. 
In both panels, the numerical results are consistent with the corresponding theoretical predictions. 
The balance of the two contributions at $l-1=2k$ explains the crossover around $l/(2k)\sim1$ from linear to quadratic growth, as illustrated in Fig.~\ref{fig:app-errg}(d).

\section{Extraction of the $kl$ term}

\begin{figure}[t!]
	\begin{center}
		\includegraphics[width=0.5\linewidth]{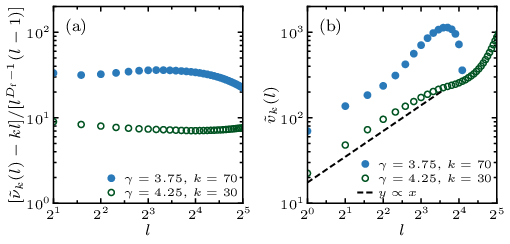}
    	\caption{
            Extraction of the linear contribution in the ball volumes of critical scale-free random networks. Filled and open circles represent the results for $(\gamma,k)=(3.75,70)$ and $(4.25,30)$, respectively.
            (a) Ratio $[\tilde{\nu}_k(l)-kl]/[l^{\df-1}(l-1)]$ used to estimate the coefficient $c$, with $\df=2.33$ and $2$ for $\gamma=3.75$ and $4.25$, respectively. The approximate plateaus identify the values of $c$ used in the scaling analysis.
            (b) Residual volume $\tilde{v}_k(l)$ after subtracting the fractal contribution. The dashed guide demonstrates the remaining degree-dependent linear term $kl$.
        }
   \label{fig:app-scalefree}
	\end{center}
\end{figure}

Since Eq.~(\ref{eq:nu_tilde}) for the ball volume contains the coefficient $c$, the value of $c$ must be determined in order to obtain the explicit form of $\tilde{\nu}_{k}(l)$.
To extract $c$ from numerical data, we consider the ratio
\begin{equation}
    c=\frac{\tilde{\nu}_k(l)-kl}{l^{\df-1}(l-1)}.
    \label{eq:coff_c}
\end{equation}
For scale-free random networks, we utilize the theoretical value of $\df$. 
Figure~\ref{fig:app-scalefree} (a) represents the right-hand side of Eq.~(\ref{eq:coff_c}) for $\gamma=3.75$ and $\gamma=4.25$ with the theoretically predicted values $\df=2.33$ and $\df=2$, respectively. 
In practice, we determine $c$ by averaging over the plateau window $[2, 2^4]$. 
The resulting values are used in the scaling analysis in Figs.~\ref{fig:1} and \ref{fig:2} of the main text and Fig. 5 (b). 
We plot the distance dependence of the residual volume $\tilde{v}_k(l)=\tilde{\nu}_k(l)-c l^{\df-1}(l-1)$ as in Fig.~\ref{fig:app-scalefree} (b), observing that the residual volume $\tilde{v}_k(l)$ increases linearly.

\begin{figure}[t!]
	\begin{center}
        \includegraphics[width=1.\linewidth]{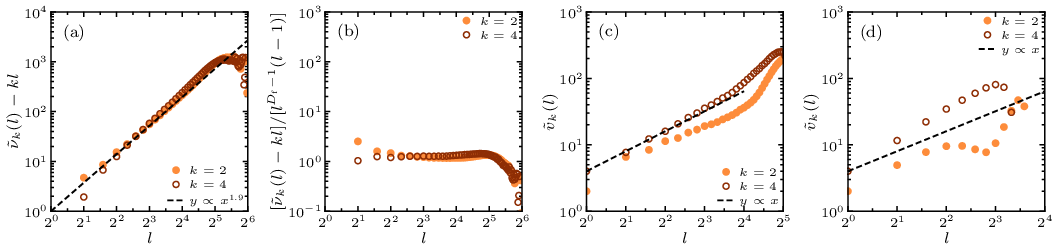}	
    	\caption{
        Extraction of the linear contribution in the ball volumes of skeletons of the protein interaction network and the protein folding network. Filled and open circles correspond to the results for root degrees $k=2$ and $4$, respectively. The results for the protein interaction network are presented in panels (a)-(c) and the results for the protein folding network are shown in panel (d).
        (a) Fractal contribution $\tilde{\nu}_k(l)-kl$ as a function of distance $l$. The dashed guide shows the fitted power $l^{1.9}$.
        (b) Ratio $[\tilde{\nu}_k(l)-kl]/[l^{\df-1}(l-1)]$ evaluated using the corresponding fitted fractal dimension $\df=1.9$; the plateau value determines the coefficient $c$.
        (c) and (d) Residual volumes $\tilde{v}_k(l)$ after subtracting the fractal contributions. The dashed lines indicate the expected linear term $kl$.
        }
   \label{fig:app-empirical}
	\end{center}
\end{figure}

By contrast, for empirical networks, we must obtain $\df$ numerically. 
To this end, we assume that the empirical networks satisfy Eq.~(\ref{eq:nu_tilde}), and estimate $\df$ from the $l$ dependence of $\tilde{\nu}_k(l)-kl$ as shown in Fig.~\ref{fig:app-empirical} (a). 
Then, using the estimated value of $\df$, we evaluate the plateau value $c$ in Fig.~\ref{fig:app-empirical} (b). 
Finally, we observe the residual volume $\tilde{v}_k(l)$ with the estimated values of $\df$ and $c$ as shown in Figs.~\ref{fig:app-empirical} (c) and (d). 
Here, panels (c) and (d) show the results for the protein interaction network and for the protein folding network, respectively.


\section{Datasets}
\subsection*{Protein-protein interaction network}
The yeast protein-protein interaction network originates from the \textit{Saccharomyces cerevisiae} data treated by Jeong \textit{et al.}~\cite{Jeong01}. The original network contains $1870$ proteins and $2240$ interactions. We treat the network as an undirected, unweighted network. After removing self-loops, the resulting network contains $1870$ nodes and $2203$ edges. We analyze its largest connected component, consisting of $1458$ proteins and $1948$ edges.
We obtained the processed dataset from KONECT~\cite{Kunegis13,KONECTMoreno}.

\subsection*{Protein-folding network}
The protein-folding network represents transitions among conformations of a $20$-residue peptide obtained from molecular-dynamics simulations~\cite{Rao04}. The original data contain $132168$ nodes and $249704$ weighted directed edges. For this analysis, we converted the data into an undirected, unweighted simple graph by removing self-loops and merging all directed edges. The resulting graph contains $132168$ nodes and $228967$ edges.
We extracted the largest connected component consisting of $132167$ nodes and $228967$ edges. The original data are available from the COSIN data archive~\cite{COSINFolding}.

\end{document}